\documentclass[prl,twocolumn,showpacs]{revtex4}
\usepackage{amsmath}
\usepackage{epsfig}

\newcommand{\nslash}{\kern 0.2 em n\kern -0.50em /}

\newcommand{\beq}{\begin{eqnarray}}
\newcommand{\eeq}{\end{eqnarray}}

\def\bq{\begin{eqnarray}}
\def\eq{\end{eqnarray}}

\def\roughly#1{\mathrel{\raise.3ex\hbox{$#1$\kern-.75em
\lower1ex\hbox{$\sim$}}}}

\begin{document}

\preprint{\hfill\parbox[b]{0.3\hsize}
{ }}

\def\bra{\langle }
\def\ket{\rangle }

\title{
Neutron orbital structure
from generalized parton distributions of $^3$He 
}
\author{M. Rinaldi\footnote{Electronic address:
matteo.rinaldi@pg.infn.it},
S. Scopetta
\footnote{Electronic
address: sergio.scopetta@pg.infn.it}}
\affiliation
{\it
Dipartimento di Fisica, Universit\`a degli Studi di Perugia, 
and INFN sezione di Perugia, via A. Pascoli 06100 Perugia, Italy
}

\begin{abstract}
 
The generalized parton distribution $H$ and $E$ 
of the $^3$He nucleus,
which could be measured
in hard exclusive processes, such as 
coherent deeply virtual
Compton scattering, are thoroughly analyzed in impulse approximation,
within the Av18 interaction.
It is found that their sum is 
dominated to a large extent by the neutron contribution:
The peculiar spin structure of $^3$He
makes this target unique for the extraction of the neutron
information.
This observation could allow to access for the first time, 
in dedicated experiments,
the orbital angular momentum of the partons in the neutron.

\end{abstract}

\maketitle

The measurement of Generalized Parton Distributions (GPDs) 
\cite{Mueller:1998fv,Radyushkin:1996nd,Ji:1996ek},
parameterizing the non-perturbative hadron structure
in hard exclusive processes, 
will be a major achievement for Hadronic Physics
in the next few years.
$H$, a target helicity-conserving quantity,
and $E$, a target helicity-flip one, 
are two of the GPDs occurring at leading twist.
Their measurement will offer novel possibilities,
such as a picture of the three-dimensional
nucleon structure
\cite{Burkardt:2000za},
and the access to the parton orbital angular momentum (OAM)
\cite{Ji:1996ek}. For the latter aim, it is mandatory to 
measure both the GPDs $H$ and $E$.
The most natural process to observe them
is Deeply Virtual Compton Scattering (DVCS),
i.e. 
$
e H \longrightarrow e' H' \gamma
$ when
$Q^2 \gg m_H^2$
(here and in the following,
$Q^2=-q \cdot q$ is the momentum transfer between the leptons $e$ and $e'$,
and $\Delta^2$ the one between the hadrons $H$ and $H'$)
\cite{Ji:1996ek,Vanderhaeghen:1999xj}.
DVCS data are being analyzed
(recent results can be found in
Refs. \cite{hermes}) and,
despite severe difficulties, GPDs are being extracted
from them
(see Refs. \cite{Guidal:2010de} and references therein). 

The issue of measuring GPDs for nuclei,
to unveil
medium modifications of bound nucleons,
has been addressed in several papers \cite{cano1,resto}.
Great attention has anyway to be paid to avoid to mistake novel
effects with
conventional ones.
To this respect, a special r\^ole
can be played by few-body nuclear targets,
for which realistic studies 
are possible and exotic effects
can be therefore distinguished. 
To this aim,
in Ref. \cite{io}, an Impulse Approximation (IA) calculation
of the flavor $q$ GPD 
of $^3$He, $H_q^3$, has been presented,
valid for $\Delta^2 \ll Q^2,M^2$,
being $M$ the nucleon mass.
The approach permits to investigate 
the coherent, no break-up channel of DVCS off $^3$He,
which
can be hardly studied at
large $\Delta^2$, due to the vanishing cross section.
It was found that
the properties of nuclear GPDs should not be trivially
inferred from those of nuclear parton distributions (PDs),
measured in 
deep inelastic scattering (DIS).

In this Letter, the approach of Ref. \cite{io} is extended 
to evaluate the GPD $E_q$ of $^3$He, $E_q^3$,
to study the possibility of accessing
the neutron information.
In facts, the properties of the free neutron 
are being investigated through experiments with nuclei, 
taking nuclear effects 
properly 
into account.
$^3$He, thanks to its particular spin structure,
is extensively used as an effective polarized neutron target
\cite{friar}.
$^3$He is therefore a serious candidate
to study the polarization properties
of the free neutron, such as its helicity-flip GPD $E_q$. 
To fully understand the importance of measuring the neutron GPDs
and the
advantages of $^3$He, 
let us first summarize the main properties of GPDs.
For a spin $1/2$ hadron target, with initial (final)
momentum and helicity $P(P')$ and $s(s')$, 
respectively, 
the GPDs $H_q(x,\xi,\Delta^2)$ and
$E_q(x,\xi,\Delta^2)$
are defined through the light cone correlator
\begin{eqnarray}
\label{eq1}
F^q_{s's}\hspace{-0.55em}& ( & \hspace{-0.55em} x,\xi,\Delta^2) =
\int {d z^- \over 4 \pi} e^{i x \bar P^+ z^-}
\bra P' s' | 
\hat O_q
| P s \ket |_{z^+=0,z_\perp = 0} 
\nonumber \\
&=& {1 \over 2 P^+} \Big [ H_q(x,\xi,\Delta^2) \bar u(P',s') 
\gamma^+ u(P,s) \Big.
\nonumber
\\
&+&
\Big . 
E_q(x,\xi,\Delta^2) \bar u(P',s') 
{i \sigma^{+ \alpha} \Delta_\alpha \over 2M} u(P,s) \Big ]~,
\end{eqnarray}
where $\hat O_q=
\bar \psi_q 
\left(- {z \over 2 } \right)
\gamma^+ \, 
\psi_q \left( {z \over 2 } \right)
$,
being $\bar P=(P+P')/2$,
$\psi_q$ the quark field, $M$ the hadron mass and
$q^\mu=(q_0,\vec q)$.
The skewedness variable, $\xi$, is defined as
$
\xi = - {\Delta^+ / (2 \bar P^+)}
$
(here and in the following, $a^{\pm}=(a^0 \pm a^3)/\sqrt{2}$).
In addition to the variables
$x,\xi$ and $\Delta^2$, GPDs depend
on the renormalization scale $Q^2$. 
This dependence, not important in this investigation, is not shown in the
following.
Among the constraints satisfied by GPDs, the ones relevant here are:
i) in the 
``forward'' limit, 
$P^\prime=P$, i.e., $\Delta^2=\xi=0$, 
DIS is recovered, and
$H_q(x,\xi,\Delta^2)$ yields the usual PD,
$
H_q(x,0,0)=q(x)
$,
while $E_q(x,0,0)$ {\it is not accessible};
ii)
the integration over $x$, yields, for
$H_q$ ($E_q$),
the contribution
of the flavour $q$ to the Dirac (Pauli)
form factor (ff) of the target:
\bq
\int_{-1}^1 
dx H_q(E_q)(x,\xi,\Delta^2) = F_{1(2)}^q(\Delta^2)~.
\eq

A fundamental result is Ji's sum rule (JSR)
\cite{Ji:1996ek}, relating
the forward limit
of the second moment of the unpolarized GPDs 
to the total angular momentum of the quark $q$
in the target, $\langle J_{q} \rangle$:
\begin{eqnarray}
\label{sumji2}
\langle J_{q} \rangle  = 
	\int_{-1}^1 
dx\, x \Big[ H_q(x,0,0) + E_q(x,0,0) \Big]~.
\end{eqnarray}

The relevance of measuring $H_q$ {\it and}
$E_q$ for the neutron through
that of $^3$He is now evident:
the combination $H_q + E_q$ is needed to study the angular momentum
content of the nucleon, through the JSR. In particular the
OAM part could be obtained from
$\langle J_{q} \rangle$,
being the helicity one measurable in DIS
and semi-inclusive DIS (SiDIS).
The relevance of this information to
understand the cumbersome spin structure
of the nucleon is apparent and,
as for any other parton observable, 
the neutron data are crucial to obtain, together with the
proton ones, a $u$ and $d$ flavor decomposition of the GPDs.
Among the light nuclei, 
$^3$He is the only one for which the nuclear combination
 $H_q + E_q$ could be dominated by
the neutron one. In facts, $^4$He is scalar and it does
not show up any $E_q$. $^2$H is also useless to this respect:
as it is easily seen in the forward limit, relevant for the
JSR, according to Eq. (2) the size of  
$E_q^A$ for a given target $A$ can be related to its ffs,
whose normalization is $F_1^A(0)=Z_A$, $F_1^A(0)+F_2^A(0)=\mu_A$, being 
$Z_A$ and $\mu_A$ the charge and the magnetic dipole moment of the target,
respectively. Using the experimental data, one gets, for $^2$H,
$F_2^2(0)\simeq - 0.14 \mu_N$, reflecting a small $E_q^2$
(in the analysis of Ref. \cite{cano1}, this contribution has
been indeed neglected).
On the contrary, in the $^3$He case, $F_2^3(0)$ is not only
sizable ($\simeq -4.13 \mu_N$), but, if summed to $F_1^3(0)$,
yields $\mu_3 \simeq -2.13 \mu_N$,
a value rather close to the neutron one, $\mu_n \simeq -1.91 \mu_N$.
As it is well known, $\mu_3$ and $\mu_n$ would be equal,
i.e., there would be no proton contribution to $\mu_3$, if $^3$He could be
described in an independent particle model with central
forces only. Although this scenario is too crude, realistic 
calculations show that the wave function lies in this configuration
with a probability close to 90 \%, a fact that
made it possible
to safely extract the neutron DIS structure functions 
from $^3$He data, as suggested in \cite{friar},
estimating carefully nuclear corrections.
In the case under investigation here, the situation is somehow different. 
GPDs are not densities, 
to be evaluated through a wave function. Anyway, this is the case
at least in the forward limit, where the JSR holds:
in that situation,
static $ ^3 $He properties can be advocated.
The aim of the present analysis is precisely to establish to what extent, 
close to the forward limit and slightly beyond it, the measured GPDs of $^3$He 
can be used to extract the neutron information and, in turn,
its OAM content. This study
is a pre-requisite for any experimental program of 
coherent DVCS  off $^3$He, a topic which is being under 
consideration at JLab. 

Let us then generalize the approach of Ref. \cite{io}, 
where the GPD 
$H_q^3$ of $^3$He has been obtained in IA.
In addition to the kinematical variables
$x$ and $\xi$, already defined,
one needs the corresponding ones for the nucleons in the target nuclei,
$x'$ and $\xi'$. 
The latter quantities can be obtained defining the ``+''
components of the momentum $k$ and $k + \Delta$ of the struck parton
before and after the interaction, with respect to
$\bar p^+ = {1 \over 2} (p + p')^+$,
being $p(p')$ the initial (final) momentum of the
interacting bound nucleon  (see \cite{io}
for details).
Using the standard procedure developed in IA studies of
DIS off nuclei \cite{fs},
the following relations
for $H_q^3,\,E_q^3$, 
in terms
of the nucleon ones, $H_q^N,\,E_q^N$,
are found
\bq
\label{H}
H_{q}^{3}(x,\xi,\Delta^2) &=& 
\sum_N
\int dE 
\int d\vec{p}
\,
\overline{\sum_{ S }}
\sum_{s }
P^N_{SS,ss}(\vec p,\vec p',E)  
\nonumber \\
& \times &
{\xi' \over \xi}
H^{N}_q(x',\xi',\Delta^2)~,
\eq
\bq
\label{HpE}
(H_{q}^{3} & + & E_q^3)(x,\xi,\Delta^2) =  
\sum_N
\int dE 
\int d\vec{p}
\nonumber \\
& \times &
(P^N_{+-,+-}(\vec p,\vec p',E)  
-
P^N_{+-,-+}(\vec p,\vec p',E))
\nonumber \\
& \times &  
{\xi' \over \xi}
(H^{N}_q+ E^N_q)(x',\xi',\Delta^2)~.
\eq
In Eqs. 
(\ref{H}) and (\ref{HpE}), 
proper components
appear of the spin-dependent
non-diagonal spectral function
of the nucleon $N$ in $^3$He
(details are given in Ref. \cite{matteo}):
\begin{eqnarray}
 \label{spectral1}
 P^N_{SS',ss'}(\vec p,\vec p',E) 
&=& 
\dfrac{1}{(2 \pi)^6} 
\dfrac{M\sqrt{ME}}{2} 
\int d\Omega _t
\\
& \times &  
\sum_{\substack{s_t}} \langle\vec{P'}S' | 
\vec{p}' s',\vec{t}s_t\rangle_N
\langle \vec{p}s,\vec{t}s_t|\vec {P}S\rangle_N~,
\nonumber 
\end{eqnarray}
where $S,S'(s,s')$ are the nuclear (nucleon) spin projections
in the initial (final) state, respectively,
and $E= E_{min} +E_R^*$, 
being $E^*_R$ the excitation energy 
of the two-body recoiling system
and 
$E_{min}=| E_{^3He}| - | E_{^2H}| = 5.5$ MeV. 
The main quantity appearing in the definition
Eq. (\ref{spectral1}) is
the intrinsic overlap integral
\bq
\langle \vec{p}s,\vec{t}s_t|\vec {P}S\rangle_N
=
\int d \vec y \, e^{i \vec p \cdot \vec y}
\bra \chi^{s}_N,
\Psi_t^{s_t}(\vec x) | \Psi_3^S(\vec x, \vec y) \ket~
\label{trueover}
\eq 
between the wave function
of $^3$He,
$\Psi_3^S$,  
with the final state, given by the
eigenfunction $\Psi_t^{s_t}$, with eigenvalue
$E = E_{min}+E_R^*$, of the state $s_t$ of the intrinsic
Hamiltonian pertaining to the system of two {\sl interacting}
nucleons with relative momentum $\vec t$, 
which can be either
a bound 
or a scattering state, and by the plane wave describing 
the nucleon $N$ in IA.

As discussed in Ref. \cite{io}, where Eq. (\ref{H}) 
has been obtained and evaluated,
the accuracy of these calculation,
since a NR spectral function
will be used to evaluate Eqs. 
(\ref{H}) and (\ref{HpE}),
is
of order 
${\cal{O}} 
\left ( {\vec p^2 / M^2},{\vec \Delta^2 / M^2} \right )$.
The interest of the present
calculation is indeed to investigate nuclear effects at low values
of $\vec \Delta^2$, for which measurements in the coherent channel 
may be performed.

\begin{figure}[t]
\includegraphics[height=9cm]{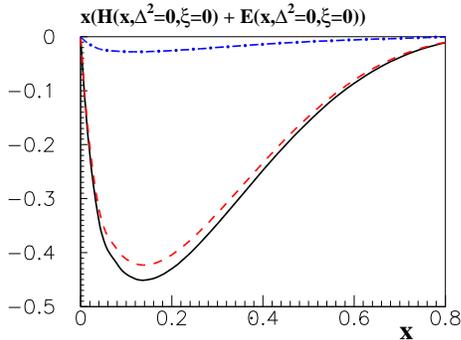}
\vskip -5cm.
\caption {
The quantity $x \sum_q (H_q^3+E_q^3)$, shown in the forward limit 
(full), together with the neutron (dashed)
and the proton (dot-dashed) contribution.
}
\label{uno}
\end{figure}

\begin{figure}[t]
\includegraphics[height=9cm]{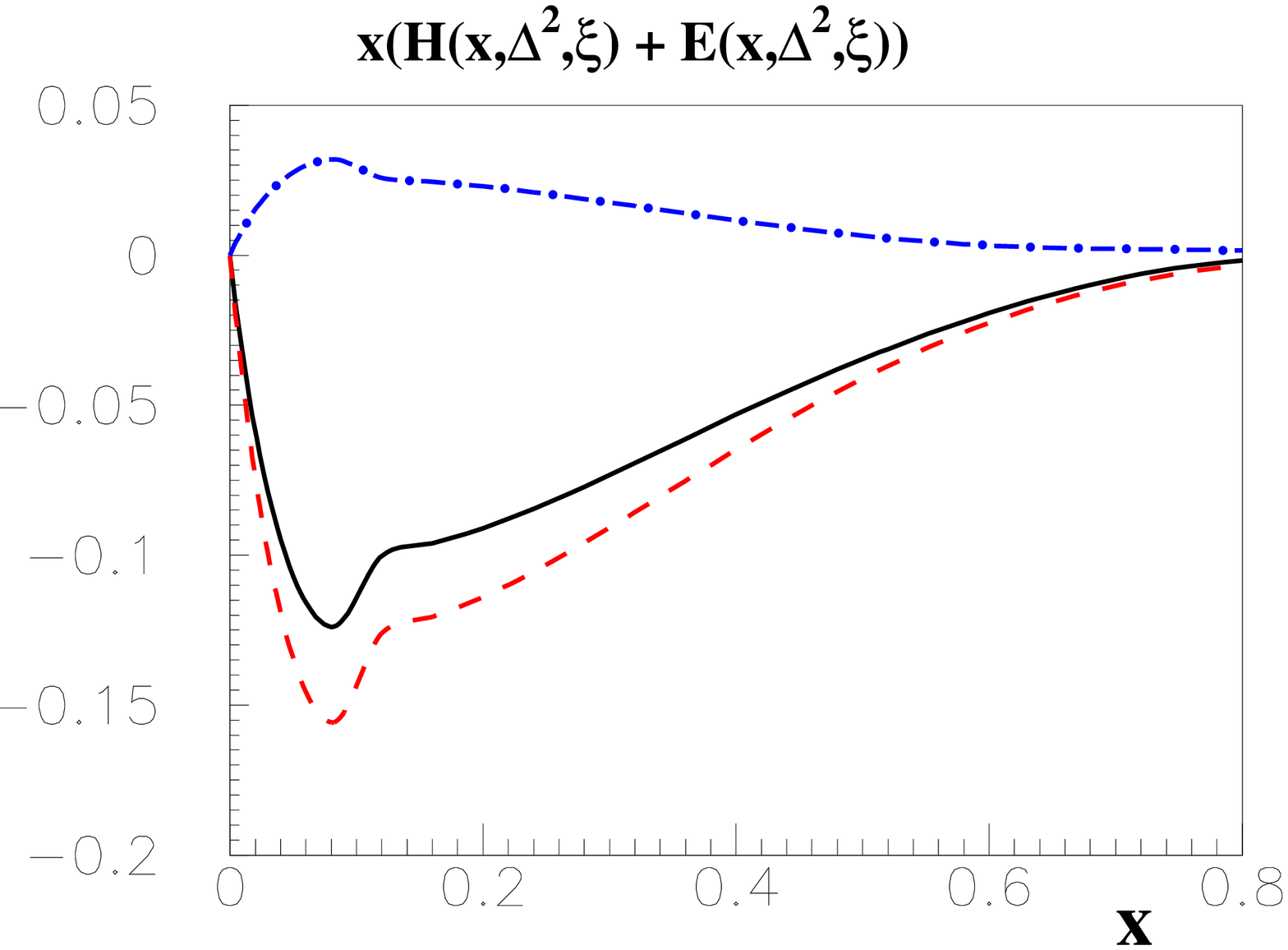}
\vskip -5cm.
\caption {
The same as in Fig. 1, but at $\Delta^2 = -0.1$ GeV$^2$
and $\xi$=0.1.
}
\label{due}
\end{figure}

\begin{figure}[t]
\includegraphics[height=9cm]{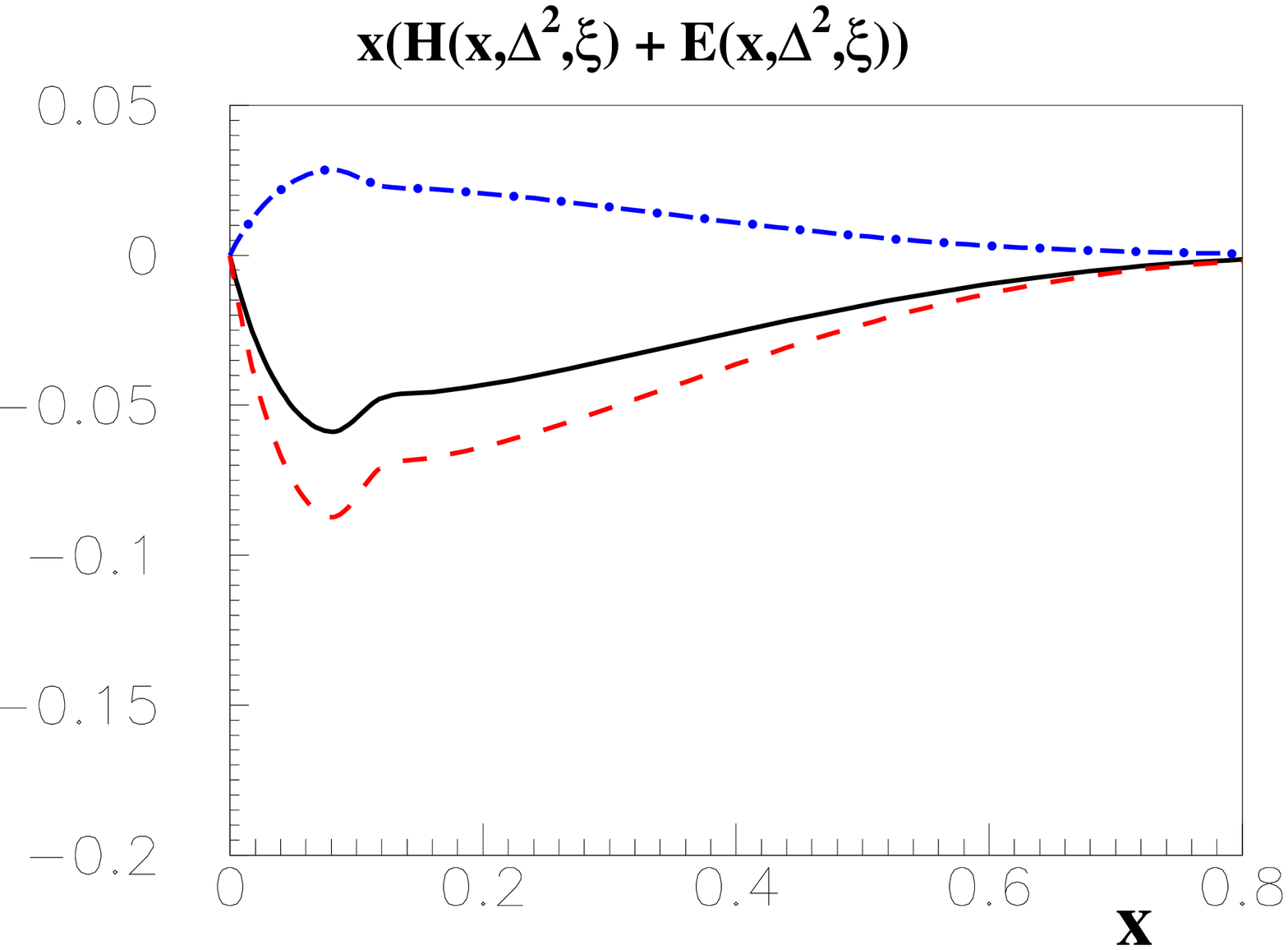}
\vskip -5cm.
\caption {
The same as in Fig. 1, but at $\Delta^2$ = -0.15 GeV$^2$,
and $\xi$=0.1.
}
\label{tref}
\end{figure}

\begin{figure}[t]
\includegraphics[height=9cm]{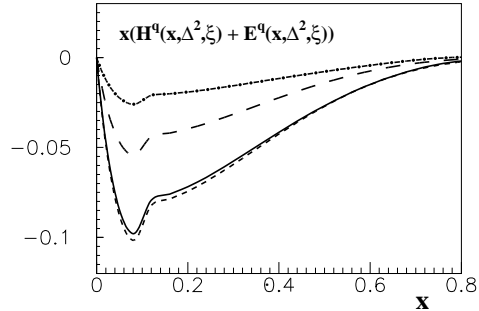}
\vskip -5cm.
\caption {
The quantity $x (H_q^3+E_q^3)$ for the $d$ (full) and $u$
(dot-dashed) flavor, at
$\Delta^2$ = -0.1 GeV$^2$,
and $\xi$=0.1. 
The neutron contributions for the $d$ (dashed) and $u$
(long-dashed) flavor are also shown.
}
\label{quattro}
\end{figure}

Eq. (\ref{HpE}),
obtained here for the first time, shows a much richer spin structure
than Eq. (\ref{H}).
Eq. (\ref{HpE})
has been evaluated in the nuclear Breit-frame,
using the exact nuclear overlaps
described above, obtained along the line of Ref. \cite{gema}, 
using the wave function \cite{tre}
corresponding to the Av18 interaction \cite{av18}.
For the nucleonic GPDs, the model of
Ref. \cite{rad1} has been used, which, despite of its
simplicity, fulfills the general properties of GPDs.
The model has been minimally extended to parametrize also the
GPD $E_q$ (see Ref. \cite{matteo} for details), assuming that
it is proportional to the 
charge of $q$ (this natural choice is used, e.g., in Refs. \cite{from}).

The only real possibility to 
establish the validity of the approach
is the comparison with experiments.
Unfortunately, data for the GPDs are not available and for $E_q^3$,
in particular, even the forward limit is unknown.
One check is in any case possible and it is therefore very
important: the quantity $H_q^3 + E_q^3$, summed over the active flavors,
can be integrated over $x$ to give the 
experimentally well-known magnetic ff of $^3$He, 
$G_M^3(\Delta^2) = F_1^3(\Delta^2) + F_2^3(\Delta^2)$
(cf. Eq. (2)).
The result we found by using this procedure \cite{matteo}
is in quantitative agreement with the Av18
one-body calculation
presented in Ref. \cite{Marcucci:1998tb}, and with
the non-relativistic part of the calculation in Ref. 
\cite{Baroncini:2008eu}.
For the values of $\Delta^2$ which are relevant for the 
coherent process under investigation here,
i.e., $-\Delta^2 \lesssim 0.2$ GeV$^2$,
our results compare well also with the data.
For higher values, the agreement is lost.
This is a well-known problem:
to get a good description of the magnetic ff of trinucleons, three-body forces 
and two-body currents have
to be introduced in the dynamical description of the
process 
(see, e.g., \cite{Marcucci:1998tb}).
If measurements were performed at high values of $-\Delta^2$,
our calculations could be improved by allowing for these effects,
a standard although lengthy procedure.  
Anyway, since coherent DVCS
cannot be measured at high $-\Delta^2$ for nuclear targets,
the good description obtained close to the static point is quite
satisfactory for the aim of the present investigation.
With the comfort of this successful check,
one can have eventually a look at the nuclear GPDs.
Results are shown in Figs. 1-4. The 
quantity $x(H_q^3+E_q^3)$ summed over the flavors $q$, which,
in the forward limit, yields
the integrand of the JSR
(cf. Eq. (\ref{sumji2})), is shown in Figs. 1-3,
in the forward limit (Fig. 1), and at
$\Delta^2 = -0.1, - 0.15$ GeV$^2$
and $\xi = 0.1$ (Figs. 2 and 3).
The shapes of the curves are very dependent 
on the nucleonic model of Ref. \cite{rad1}, used
as input in the calculation, but
one should not forget that
the aim of this analysis, for the moment being,
is that of getting a clear estimate of the 
proton and neutron contribution to the nuclear observable,
a feature rather independent on the nucleonic model.
The difference in size of the curves in Figs. 1 to 3  
reflects the dramatic effect of increasing $\Delta^2$,
a feature basically governed by the ff.
The most evident and interesting result is actually
that the contribution of the neutron is impressively
dominating the nuclear GPD at low $\Delta^2$, with the proton
contribution growing fast with increasing $\Delta^2$,
ranging from a few percent in the forward limit, to
15 \% at most at $\Delta^2 = -0.1$ GeV$^2$ but being already
30 \% at $\Delta^2 = - 0.15$ GeV$^2$.
On the contrary, as shown in Fig. 4, for the flavor $d$ the impressive
dominance of the neutron contribution varies slowly with increasing 
$\Delta^2$. All these features can be understood qualitatively
looking at Eq.(\ref{HpE}) which, being rather involved, can
be usefully sketched as follows
\bq
H^3_q+E^3_q \approx 
P^{3}_p \otimes (H^p_q+E^p_q) + P^{3}_n \otimes (H^n_q+E^n_q)~,
\eq
\noindent
where $ P^3_{p(n)} $ describes the proton (neutron) 
dynamics in $ ^3 $He, while $  (H^{p(n)}_q+E^{p(n)}_q) $ is the 
contribution of the flavor $ q $ to the GPDs of the proton (neutron).
As already explained, due to the 
spin structure of
$ ^3 $He, $ P^3_n $ 
is quite larger than $ P^3_p $, 
justifying the relevance of the neutron contributions in Fig. 1.
Anyway, with increasing $\Delta^2$, for the $u$ flavor,
the term $ H^p_u+E^p_u $ gets much larger
than $ H^n_u+E^n_u $, explaining the growth with $\Delta^2$
of the relative size of the proton contribution with respect to the 
neutron one, shown in Figs. 2-4.
This does not occur for the $d$ flavor, 
and the dominance of the neutron contribution is not
hindered by increasing $\Delta^2$ as for the $u$ flavor (cf. Fig. 4).
This happens also because  
one half of the $d$
content of $^3$He comes from the neutron,
while only one fifth of the $u$ one comes from it.
In any case, the fact that
the proton contribution gets sizable
going towards less forward situations should
not hinder the extraction of the neutron properties
close to
the forward limit, where the most important information,
related to the OAM of the partons
in $^3$He and then in the neutron, is expected.

A comment is in order concerning the possibility of accessing
neutron GPDs in {\it incoherent} DVCS off the neutron in nuclear targets,
i.e., the process when the interacting neutron is detected
together with the scattered electron and the produced
photon. An experiment of this type has been approved
at the 12 GeV program of JLab \cite{silvia} for a $^2$H target.
Although these kind of processes are hindered
by Final State Interactions of the detected neutron, 
important information, complementary
to that obtained with the coherent process proposed here,
will be collected. In the near future, we plan therefore to investigate
also incoherent DVCS off the neutron in $^3$He.

In this work, a calculation of the GPDs $H_q,E_q$ of 
$^3$He has been presented,
proposing coherent DVCS off $^3$He at low $\Delta^2$
as a key-process to obtain the neutron information. 
The directions to improve the treatment
are clear. In particular,
two-body currents and three-body
forces can be included into the game, if high values of $\Delta^2$ will be 
experimentally studied.
At the same time, a Light-Front analysis of the process,
already started in SiDIS \cite{alessio}, can be performed, to have, from the
beginning, a relativistic framework for the investigation.

It is a pleasure to thank L.P. Kaptari and G. Salm\`e for
enlightening discussions.

\end{document}